\documentclass[natbib, graybox]{svmult}

\usepackage{mathptmx}       
\usepackage{helvet}         
\usepackage{courier}        
\usepackage{type1cm}        
\usepackage{makeidx}         
\usepackage{graphicx}        
\usepackage{multicol}        
\usepackage[bottom]{footmisc}

\usepackage{amssymb,amsmath}
\usepackage{hyperref}

\makeindex             

\newcommand{\beq}{\begin{equation}}
\newcommand{\eeq}{\end{equation}}

\newcommand{\lsim}{\ \raise
-2.truept\hbox{\rlap{\hbox{$\sim$}}\raise5.truept\hbox{$<$}\ }}
\newcommand{\gsim}{\ \raise
-2.truept\hbox{\rlap{\hbox{$\sim$}}\raise5.truept\hbox{$>$}\ }}
\newcommand{\simsim}{\ \raise
-2.truept\hbox{\rlap{\hbox{$\sim$}}\raise5.truept\hbox{$\sim$}\ }}

\def\Q{\ifmmode\mathcal{Q}\else$\mathcal{Q}$\fi}

\begin{document}

\title*{The Panchromatic Hubble Andromeda Treasury. Progression of Large-Scale Star Formation across Space and Time in M\,31}
\titlerunning{Large-Scale Star Formation across Space and Time in M\,31 with PHAT} 

\author{\bf Dimitrios A. Gouliermis,
Lori C. Beerman,
Luciana Bianchi,
Julianne J. Dalcanton,
Andrew E. Dolphin, 
Morgan Fouesneau, 
Karl D. Gordon,
Puragra Guhathakurta,
Jason Kalirai,
Dustin Lang,
Anil Seth,
Evan Skillman,
Daniel R. Weisz, 
and Benjamin F. Williams
}

\authorrunning{D. A. Gouliermis et al.} 

\institute{
Dimitrios A. Gouliermis 
\at University of Heidelberg, Centre for Astronomy, Institute for Theoretical Astrophysics, Albert-Ueberle-Str.\,2, 69120 Heidelberg, Germany;  \\
Max Planck Institute for Astronomy, K\"{o}nigstuhl 17, 69117 Heidelberg, Germany\\
email: dgoulierm@googlemail.com; dgoulier@mpia.de 
}
\maketitle


\abstract{We investigate the clustering of early-type stars younger than 300\,Myr on galactic scales in M\,31. 
Based on the stellar photometric catalogs of the Panchromatic Hubble Andromeda Treasury program that also 
provides stellar parameters derived from the individual energy distributions, our analysis focused on  the young 
stars in three star-forming regions, located at galactocentric distances of about 5, 10, and 15\,kpc, corresponding 
to the inner spiral arms, the ring structure, and the outer arm, respectively.  We apply the two-point correlation function 
to our selected sample to investigate the clustering behavior of these stars across different time- and length-scales. We 
find that young stellar structure survives across the whole extent of M\,31 longer than 300\,Myr. Stellar distribution in all 
regions appears to be self-similar, with younger stars being systematically more strongly clustered than the older, which 
are more dispersed. The observed clustering is interpreted as being induced by turbulence, the driving source for which 
is probably gravitational instabilities driven by the spiral arms, which are stronger closer to the galactic centre.
}

\section{Introduction}
\label{sec:intro}

\begin{figure}[t]
\sidecaption
\includegraphics[width=0.64\textwidth]{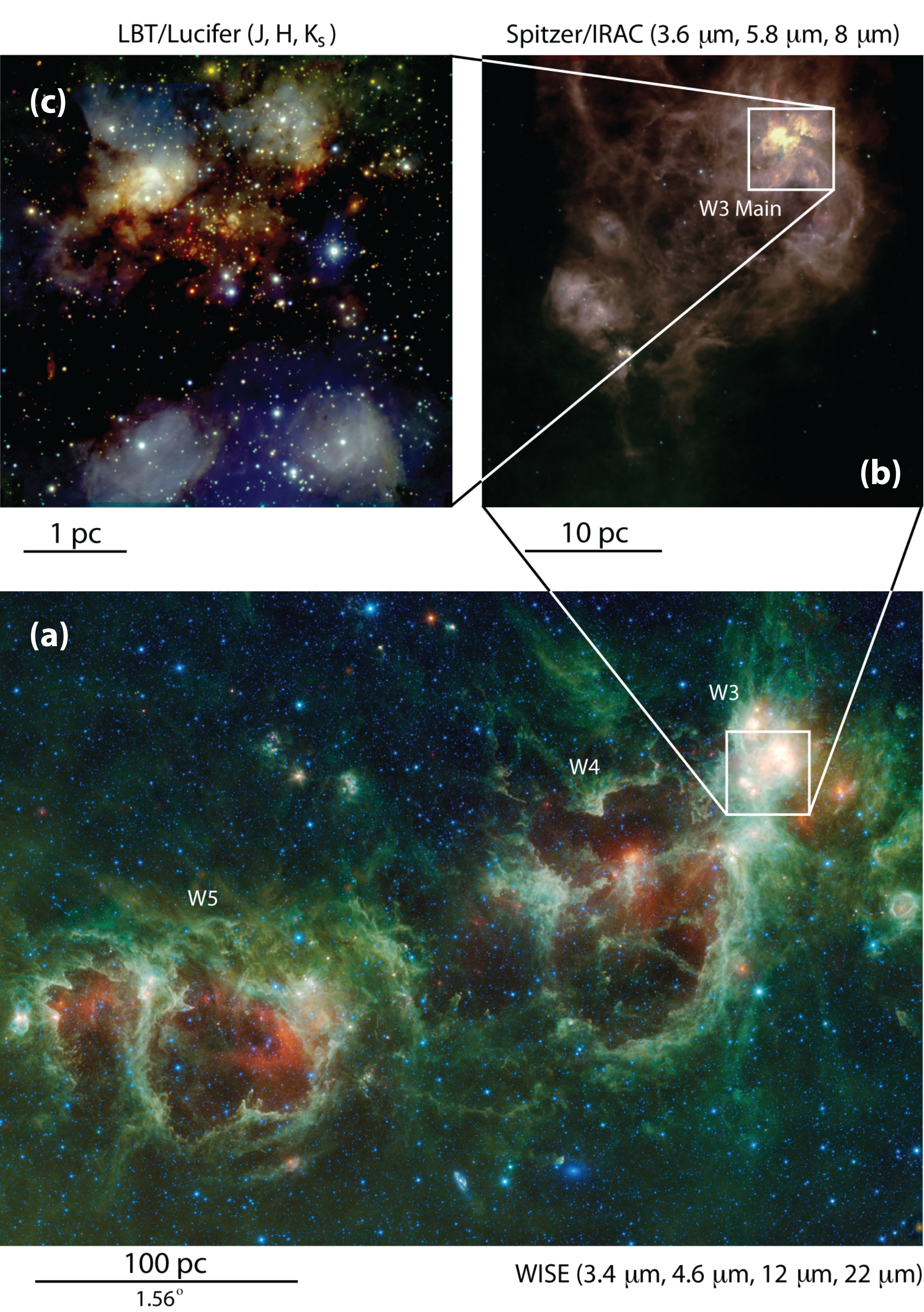}
%
%
\caption{Hierarchy in the ISM and stellar clustering in the general area of W3 complex. 
(a) On 100\,pc-scale, the nebular structures W4 and W5 are shown in the mid-IR image from NASA/WISE, 
which covers the general region of the OB association Cas~OB6. (b) On 10\,pc-scale, the 
star-forming region W3, considered as part of W4, is shown from NASA/Spitzer images, with the active star-forming 
region W3\,Main showing bright PAH emission at 8$\mu$m. (c) Finally, on the 1~pc-scale 
high-resolution near-IR images from {\sc Luci} camera on the LBT reveal the embedded star-forming cluster of W3\,Main
\citep{bik12,bik14}. WISE image credit: NASA/JPL-Caltech/WISE Team.}
\label{fig:w3example}       
\end{figure}

Stars are born in groups \citep[][]{lada03} of various sizes and with various degrees of gravitational 
self-binding \citep[][]{elmegreen-ppiv}. Observed length-scales of star formation range from few 100-pc, 
typical for loose stellar complexes, to few 10-pc, characteristic of unbound stellar aggregates and OB 
associations, and finally down to few pc, indicative of more compact young star clusters. All these types 
of stellar clusterings are not independent from each other and instead, they are structured into a 
hierarchical fashion (\citealt{E2011EAS31}), similar to that of the interstellar matter (ISM). Giant molecular 
clouds are indeed hierarchical structures \citep{elmegreen96, stutzki98}, indicating that scale-free processes  
determine their global morphology, turbulence being considered the dominant \citep{elmegreenscalo04, 
maclowklessen04}. It stands to reason that  mechanisms regulating star formation \citep{mckee+ostriker07} 
consequently shape stellar structures, which then build up in a hierarchical fashion, due to 
the self-similar nature of turbulent cascade \citep[e.g.,][]{klessen00}. This structuring behavior is 
exemplified  in Fig.\,\ref{fig:w3example} for the Galactic star-forming complex W3/4/5.

The formation of stars proceeds hierarchically also in time. The duration of star formation tends to increase 
with the size of the region as the crossing time for turbulent motions  \citep{efremelme98}.  Small regions form stars quickly and large 
regions, which contain the small ones, form stars over a longer period. This relation between length- and time-scale underscores 
the perception that both, cloud and stellar structures, come from interstellar gas turbulence and suggests that star formation in a 
molecular cloud is  finished within only few turbulent crossing times \citep[e.g.,][]{ballesteros-paredes99, elmegreen00,
hartmann01}.  

While this picture of clustered star formation explains the formation of young stellar assembles across different scales, the 
relation between ISM structure and stellar clustering is not well understood. In addition, apart from star formation, environmental 
conditions (local feedback, galactic dynamics, etc) influence the morphology of stellar clustering, and the observed variety of stellar 
systems in size, shape, and compactness. Dissentangling the relative importance of the heredity of star formation to the stellar clustering 
(``nature'') in comparison to the environmental influence on its morphology and survival (``nurture'') plays, thus, an important  role
to our understanding of clustered star formation.

We present our investigation of stellar structures formation over galactic scales from the census of bright blue stars, distributed over the typical 
length-scale of spiral arms. Our dataset is observed with the most complete stellar survey ever 
performed of M\,31, obtained by the {\sl Panchromatic Hubble Space Telescope Andromeda Galaxy Treasury} 
program\footnote{\href{http://www.astro.washington.edu/groups/phat/Home.html}{http://www.astro.washington.edu/groups/phat/Home.html}}. 
We describe our analysis and present first results on the clustering behaviour of young stars up to ages of $\sim$\,300\,Myr in  this  
galaxy.

\begin{figure}[t]
\sidecaption
\includegraphics[width=\textwidth]{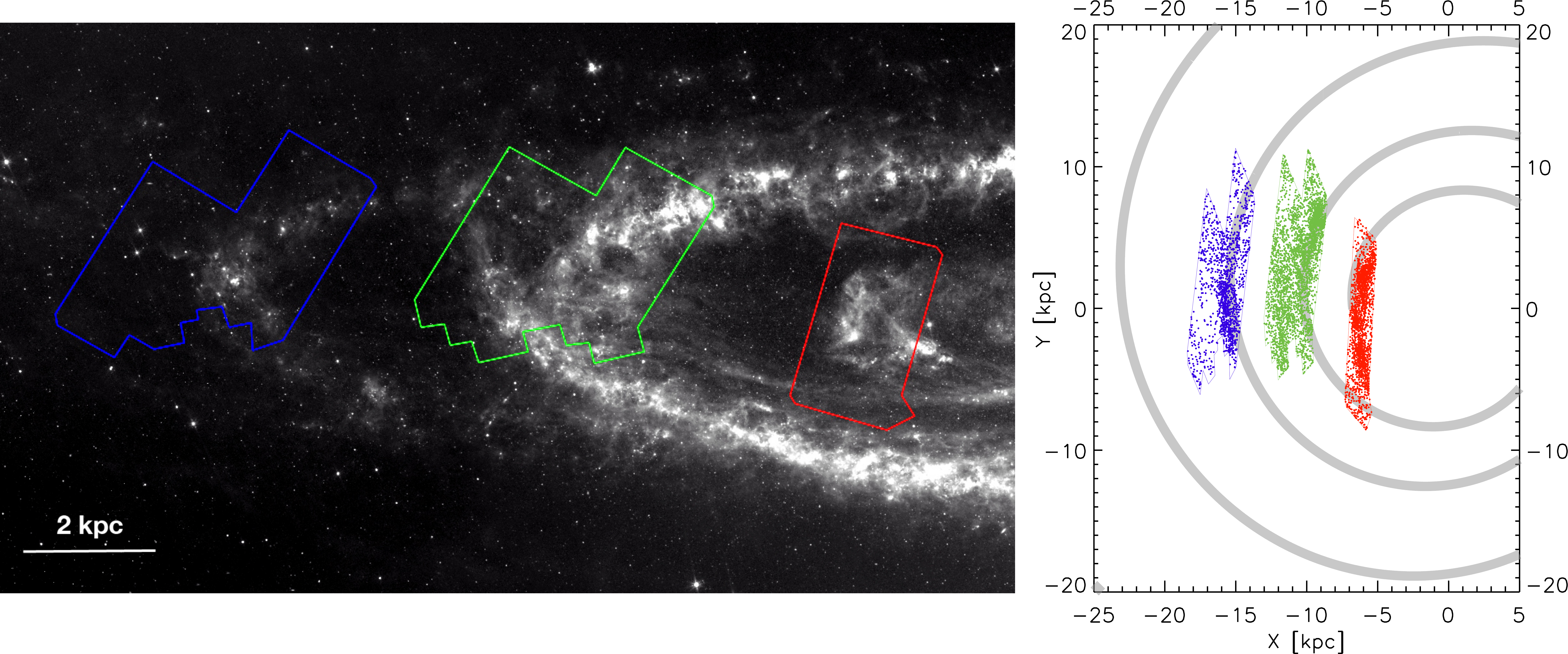}
%
%
\caption{Footprints of the selected regions on the {\sl Spitzer} 8$\mu$m image of M\,31 (left panel),
and coverage of the selected regions in respect to the Andromeda disk corrected for the inclination 
of the galaxy, shown face-on (right panel), in respect to the spiral arms in accordance to \cite{arp64}. 
Area\,9 (red) covers the northeastern turn-over 
of the inner spiral arm at $\sim$\,5\,kpc from the centre, and area\,15 (green) that of the second arm 
on the 10\,kpc star-forming ring of Andromeda. Area\,21 (blue) located at distance $\sim$\,15\,kpc 
from the centre, includes various star-forming regions at the outskirts of M\,31.}
\label{fig:footprints}       
\end{figure}

\section{Observational material}
\label{sec:data}

The {\sl Panchromatic Hubble Space Telescope Andromeda Treasury}\index{Panchromatic 
Hubble Space Telescope Andromeda Treasury} (PHAT)\index{PHAT} 
program \citep{dalcanton12} provides deep coverage of 1/3 of M\,31
galaxy in six filters with {\sl HST}. The survey spans the north east quadrant 
of the galaxy, continuously imaging from the nucleus to the last obvious regions  
of star formation visible with GALEX \citep[e.g.,][]{thilker05}.
This part of M\,31 is selected because of its lowest internal extinction, the highest intensity 
regions of unobscured star formation, and the least contamination from M\,32. 
Imaging is performed blue-ward of 4000\AA, in filters F275W and F336W with  
the WFC3 camera, in the optical F475W and F814W filters with ACS/WFC, and in 
the near-IR, in the WFC3 filters F110W and F160W \citep{williams14}. This 
wide panchromatic coverage baseline allows us to confidently estimate  stellar effective  
temperatures, masses, ages and reddenings through a self-consistent Bayesian Spectral Energy 
Distribution (SED) fitting technique (Gordon et al., in prep), 
and identify specific features on the color-magnitude diagrams (CMDs) for hot and cool stars for a wide 
range of extinctions.  The homogeneity of the stellar photometric catalogs produced by PHAT over a wide 
spatial coverage provides the unique opportunity to address the clustering behavior of star 
formation at length-scales of few kpc, corresponding to spiral arms, down to the few pc scale, 
where individual star clusters reside \citep[][]{johnson12, fouesneau14, simones14}. 

Observations of the PHAT survey were performed in 3\,$\times$\,6 mosaics of 18 parallel ACS and WFC3 pointings. Each 
mosaic builds a so-called `brick', roughly corresponding to regions of $1.5\times3$\,kpc at the distance of M\,31. In total, 23 
such bricks tile the complete survey area. The regions of interest in our study are covered by few such bricks, all including 
portions of the star-forming spiral arms and 10-kpc ring of Andromeda (Fig.\,\ref{fig:footprints}). We name each area after the identification 
number of its first brick, e.g., areas 9, 15 and 21. Area 9 covers only one brick, while areas 15 and 21 cover somewhat more than 
two adjacent bricks. These (apart from a brick that covers part of the bulge) are the only areas for 
which preliminary stellar parameters are available through our SED fitting technique. We distinguish the bright hot 
stars with the best-fit values in each area, based on three selection criteria: 
\begin{enumerate}
\item Stars  with $T_{\rm eff} \geq 10,000$~K, i.e., spectral type earlier than $\sim$~A0V.
\item Stars detected in at least three of the bluest filters, i.e., in F275W (NUV), F336W (U),  and F475W (B).
\item Stars with equivalent U-band magnitude $m_{\rm 336} \leq 25.25$.
\end{enumerate}
With these criteria, we ensure to select the brightest young stars with $L \gsim 80$\,L$_\odot$, and also the most 
accurate determinations of physical parameters based on the preliminary assumed stellar models.

\section{Evolution of young stellar structure in M\,31}
\label{sec:clustering}

\subsection{The two-point correlation function}
\label{sec:tpcf}
The spatial distribution of the early-type stars in the selected areas is characterized in terms of 
the two-point correlation function (TPCF)\index{Two-point correlation function}. Originally introduced 
for measuring cosmological structure \cite[e.g.,][]{peebles80}, the 
TPCF\index{TPCF} is a robust measurement of the degree of clustering in a sample of sources. 
Based on the original estimator by \cite{peebleshauser74} the TPCF is estimated by counting the 
pairs of sources with different separations $r$. This function is defined as \citep[see, e.g.,][]{scheepmaker09}:
\begin{equation} 
1+\xi (r) =  \frac{1}{\bar{n}N}\sum_{i=1}^{N} n_{i}(r),
\label{eq:tpcf}
\end{equation} 
which measures the surface density enhancement $n_{i} (r)$ within radius 
$r$ from star $i$ with respect to the global average surface density $\bar{n}$ of the 
total sample of $N$ stars. For a fractal distribution the TPCF yields a power-law dependency 
with radius of the form $1+\xi (r) \propto r^{\eta}$.  By definition, the total number of stars $N_r$  
within an aperture of radius $r$ will be then increasing as $N_r\propto r^{\eta}\cdot r^{2}  = r^{\eta+2}$. 
The power-law index $\eta$ is thus related  to the (two-dimensional) fractal dimension, $D_2$, of the 
(projected) distribution as $D_{2} = \eta +2$ \citep{mandelbrot83}. 

\begin{figure}[t]
\sidecaption
\includegraphics[width=\textwidth]{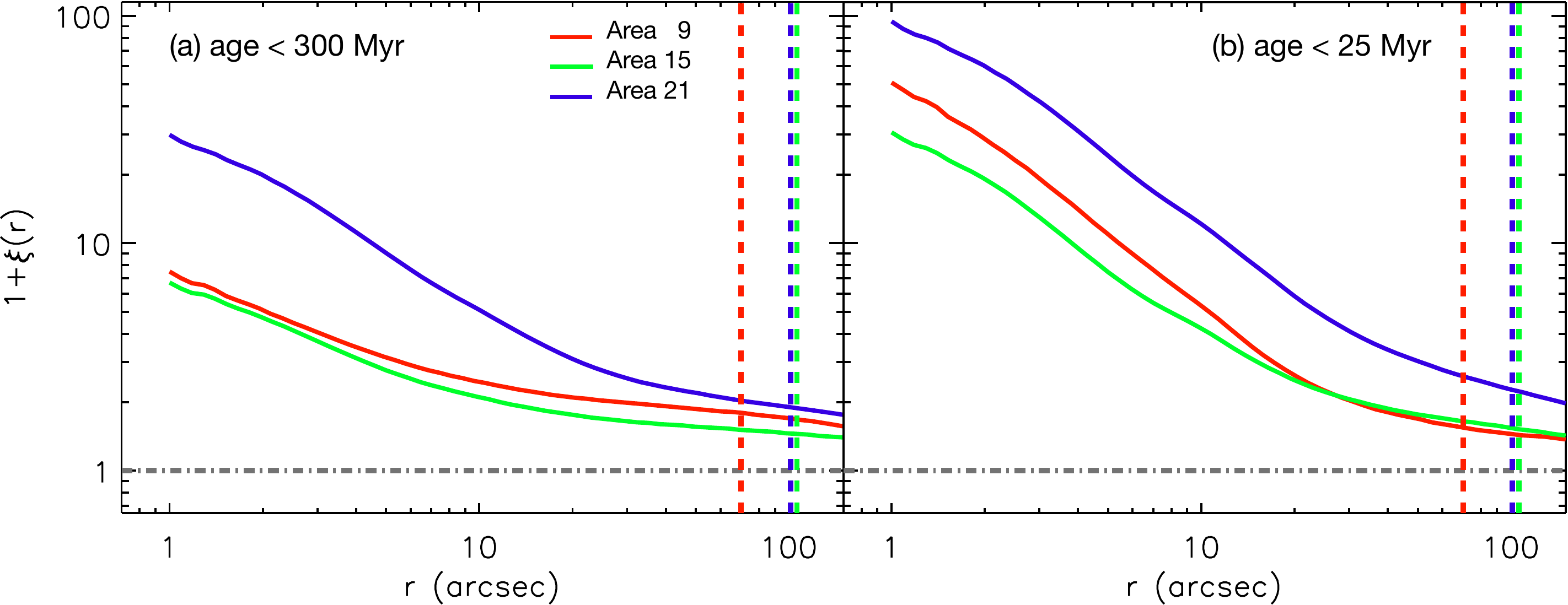}
%
%
\caption{The TPCF of the three areas of interest constructed for stars in two different age ranges, i.e., 
the complete samples with stellar ages up to $\sim$\,300\,Myr and a subsample of stars with ages 
up to 25\,Myr. All TPCFs behave as power-laws up to specific separations, beyond which the TPCF index
evidently changes. Vertical lines, coloured accordingly, correspond to the limiting separations beyond which 
the effects of the limited field-of-view become dominant, and thus TPCF measurements are not trustful \citep[see, e.g.,][]{gouliermis14}.}
\label{fig:tpcf_ex}       
\end{figure}

This TPCF formalism allows for the direct interpretation of the absolute value of $1+\xi(r)$, without any comparison with 
a reference random distribution \citep[as, e.g., proposed by][]{gomez93}, because for a random (Poisson) stellar distribution  
the value of the TPCF is always $1+\xi(r) = 1$, independent of $r$. For 
truly clustered distributions the value of the TPCF is a measurement of the clustering degree of the stars, and is always $> 1$;  higher
value at a given separation, stands for higher degree of clustering in the sample \citep[for a complete description see][and references therein]{gouliermis14}.
The behaviour of the TPCF is illustrated by the examples given in Fig.\,\ref{fig:tpcf_ex}, where the TPCFs of the three 
areas for the complete sample of stars with ages up to $\sim$\,300\,Myr and stars with ages up to $\sim$\,25\,Myr, are shown. 

The log-log plots of Fig.\,\ref{fig:tpcf_ex} show that all TPCFs have broken power-law shape. We determine the index $\eta$ of each part 
of the TPCF by applying a Levenberg--Marquard nonlinear least square minimization fit \citep{levenberg44, marquardt63}. The fitting function
has the form:
\begin{eqnarray}
\log{\left(1+\xi({r})\right)} = \left\{
          \begin{array}{r@{\,\,}l}
          \alpha + \beta \cdot \log{(r)} & ~~~~~{\rm for}\, \log{(r)} < \delta\\
          & \\
          \alpha + \left( \beta - \gamma \right) \cdot \delta + \gamma \cdot \log{(r)} & ~~~~~{\rm for}\, \log{(r)} > \delta\\
          \end{array}\right.
\label{eq:funcform}
\end{eqnarray}
where $\beta$ and $\gamma$ are the power-law slopes and $\delta$ is the logarithm of the 
position of the separation break along the abscissa, $S_{\rm break}$, where the TPCF index changes. Both slopes and 
the separation break are free parameters in our fit.

From the comparison 
of the TPCFs of Fig.\,\ref{fig:tpcf_ex} we derive two interesting results. First, the younger stars (age $\lsim$\,25\,Myr) show a more fractal,
i.e., clumpy, clustering than stars in the whole sample (age \lsim\,300\,Myr). This is demonstrated by both the higher values of $1+\xi$ at small separations 
and the steeper slopes of their TPCFs (the latter only for areas\,9 and 15). Second, while the TPCF index of areas\,9 and 15 changes significantly, 
becoming steeper for the younger sample, that of area\,21 remains unchanged between the two samples. This indicates that 
older populations (up to $\sim$\,300\,Myr) in areas 9 and 15 have higher filling factors (flatter TPCF slopes) than those in area\,21. Considering 
that the latter is the most remote  area, away from the centre of M\,31, we can only assume that the more dispersed distribution
of older stars in areas 9 and 15 is due to the dynamics of the galaxy. We investigate in more detail the evolution of the TPCF with stellar age
in the following sections.

\subsection{Evolution of the TPCF with time}
\label{sec:tpcf_evol}

We construct the TPCF of the young stellar samples in the three areas of interest for stars of different ages. We divide each
sample to several subsamples of stars, according to the best-fit ages assigned to each star with the SED-fitting.
These ages are constrained by the stellar evolution models \citep{girardi10} and atmosphere templates \citep{castellikurucz} 
preliminary used in our technique. Therefore, we select our subsamples 
of stars according to their  effective temperatures, which are somewhat better-constrained by the 
SED fitting, and are clear indicators of the stellar evolutionary stage. In order for our analysis to be performed 
in samples of equivalent statistical significance (i.e., not being affected by different number statistics) we divided 
each stellar catalog into $T_{\rm eff}$-determined subsamples, containing the same number of stars, corresponding to $\sim$\,10\% 
of the total. 
We divided thus each stellar catalog into nine (in the case of area 9 into ten) subsamples that contain almost identical number 
of stars at different evolutionary stages. This number is $\sim$\,5,000 in areas\,9 and 21 and $\sim$10,000 in area\,15. An example of 
the subsamples and the derived TPCF parameters for area\,15  is shown in Table\,\ref{tab:a15samples}.

\begin{figure}[t]
\sidecaption
\includegraphics[width=\textwidth]{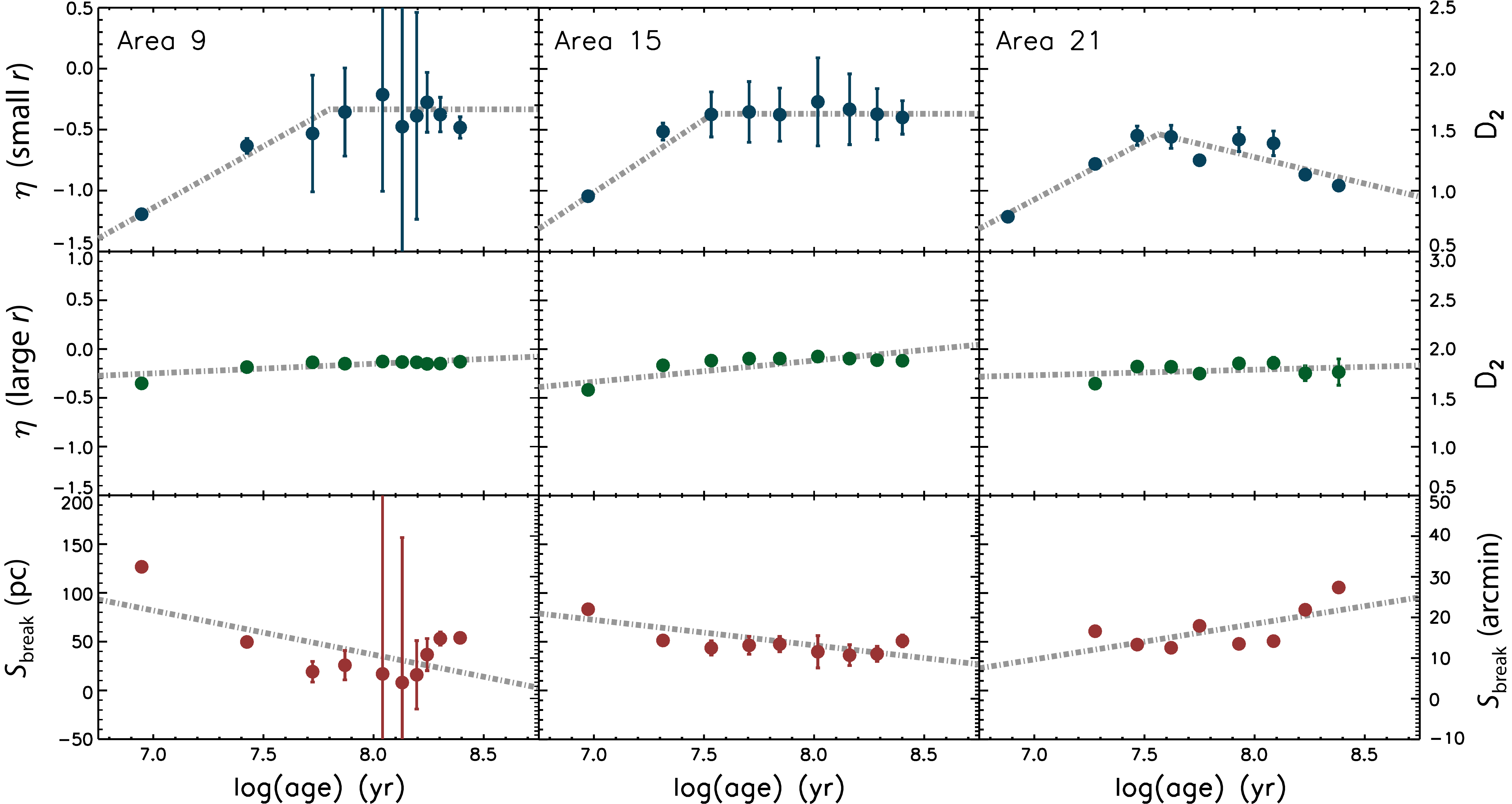}
%
%
\caption{Evolution of the TPCF as a function of stellar age for the three areas of interest. {\em Top panel}: 
Relation of the TPCF index $\eta$ (or the equivalent fractal dimension $D_2$, given on the right ordinate) to average stellar age 
for small separations (smaller than the length $S_{\rm break}$ where the TPCF slope changes). {\em Middle panel}: 
The same relation for large separations, i.e., $r \geq S_{\rm break}$. {\em Bottom panel}: Relation of  
$S_{\rm break}$  to average stellar age for each subsample. The grey dash-dotted lines are linear fits to the data.
The TPCF of stars in the youngest age range of area\,21 is a single power-law, and therefore there is no measurement for
$D_2$ at large separations and $S_{\rm break}$ for this age.}
\label{fig:tpcf_evol}       
\end{figure}

We determine the TPCF index $\eta$ for both small and large separations, as well as the separation $S_{\rm break}$  where $\eta$ 
changes, for each subsample by applying again a Levenberg--Marquard nonlinear least square minimization fit. The results are given 
in Fig.\,\ref{fig:tpcf_evol}, where we show the relation of index $\eta$ (and the derived $D_2$) to the average age of the 
corresponding stellar subsample for small ($r \leq S_{\rm break}$; top panel) and large ($r \geq S_{\rm break}$; middle panel) separations. 

\begin{table}[t]
\caption{Selected stellar subsamples and derived TPCF parameters for area\,15. }
\label{tab:a15samples} 
\begin{tabular}{p{1.10cm}p{1.25cm}p{1.75cm}p{1.05cm}p{2cm}p{2.cm}p{1.75cm}}
\hline\noalign{\smallskip}
$\langle \log{T_{\rm eff}} \rangle^a$ & $\langle {\rm age} \rangle^b$ & age limits$^{c}$ & $\bar{n}^{d}$ ($10^{-4}$& $D^{e}$ (small & $D$ (large & Separation\\   
 & (Myr) & (Myr) & pc$^{-2})$&separations)& separations)& break$^{f}$  (pc) \\   
\noalign{\smallskip}\svhline\noalign{\smallskip}
       4.38    &            ~~~~9   $\pm$           ~~7    &            ~~0.2   $-$          ~~29.0    &     2.7&     0.955   $\pm$         0.004    &          1.583   $\pm$         0.012    &          83.37   $\pm$          ~~3.83    \\     
      4.26    &           ~~21   $\pm$          15    &            ~~1.0   $-$          ~~52.7    &        3.3    &  1.485   $\pm$         0.069    &          1.835   $\pm$         0.016    &          51.40   $\pm$          ~~4.71    \\
      4.21    &           ~~34   $\pm$          24    &            ~~0.9   $-$          ~~75.4    &         3.5    &  1.626   $\pm$         0.185    &          1.883   $\pm$         0.020    &          43.70   $\pm$          ~~7.27    \\
      4.17    &           ~~51   $\pm$          30    &            ~~1.0   $-$          ~~84.1    &          3.3  & 1.646   $\pm$         0.249    &          1.904   $\pm$         0.028    &          46.37   $\pm$          ~~9.14    \\
       4.14    &           ~~70   $\pm$          37    &            ~~3.8   $-$         109.0    &           2.7   &  1.624   $\pm$         0.218    &          1.904   $\pm$         0.026    &          47.61   $\pm$          ~~7.85    \\
     4.11    &          104   $\pm$          40    &            ~~7.7   $-$         151.0    &              2.2     &   1.729   $\pm$         0.361    &          1.925   $\pm$         0.024    &          39.79   $\pm$         16.49    \\
     4.08    &          145   $\pm$          44    &           10.2   $-$         188.0    &                 2.6     &   1.668   $\pm$         0.290    &          1.904   $\pm$         0.021    &          36.28   $\pm$         10.70    \\
       4.04    &          193   $\pm$          51    &           11.0   $-$         242.0    &                2.5    &  1.628   $\pm$         0.209    &          1.888   $\pm$         0.020    &          37.58   $\pm$          ~~7.69    \\
4.01    &          252   $\pm$          63    &           13.1   $-$         318.0    &                       2.6    & 1.601   $\pm$         0.137    &          1.882   $\pm$         0.022    &          50.90   $\pm$          ~~6.06    \\
\noalign{\smallskip}\hline\noalign{\smallskip}
\end{tabular}
$^a$ Average logarithmic effective temperature of stars in the subsample.\\
$^b$ Average age of stars in the subsample. \\
$^c$ Age limits of stars in the subsample. \\
$^d$ Mean stellar surface density. \\
$^e$ Two-dimensional fractal dimension derived from the TPCF index $\eta$ as $D=\eta+2$. \\
$^f$ Separation limit, where the TPCF shows a significant change in its slope $\eta$.
\end{table}

For small separations the plots of $\eta$ (or $D_2$) versus stellar age show that there is a clear time evolution of the TPCF, with $D_2$ 
being smaller at younger ages, and  becoming larger for older ages.  In all areas $D_2$ starts with values \lsim\,1 at the smaller age bin 
of $\langle$age$\rangle \sim$\,9\,Myr, and increases almost monotonically to a value of $\sim 1.6$ at $\langle$age$\rangle \sim$\,75\,Myr 
for area\,9 and 35\,Myr for area\,15 (see also Table\,\ref{tab:a15samples}), and to $\sim$\,1.4 at $\langle$age$\rangle \sim$\,45\,Myr for area\,21. 
It is worth noting that while for older stars in areas\, 9 and 15 the TPCF slope `stabilizes' at almost constant value through the whole remaining age extent,
in area\,21 it drops again to the value of $D_2 \sim$\,1. For separations $r \geq S_{\rm break}$ (Fig.\,\ref{fig:tpcf_evol}, middle panel)
the TPCF shows no important evolution with stellar age, having a value of $D_2$ $\sim 1.85 \pm 0.02$ (area\,9), $\sim 1.85 \pm 0.11$ (area\, 15),
and $\sim 1.79 \pm 0.07$ (area\,21). While this value for area\,21 is somewhat smaller than those in areas\,9 and 15, all values 
are high, close to the geometrical (projected) dimension, and thus consistent with almost uniform (non-clustered) stellar distributions. 
Interestingly, we observe from the bottom panels of Fig.\,\ref{fig:tpcf_evol},  that the limit of the fractal regime, $S_{\rm break}$, seems to also  
evolve with stellar age. 

\section{Concluding Remarks and Summary}

The value $D_2 \lsim 1$ found for the youngest stars at separations $r \leq S_{\rm break}$ indicates that 
star formation is clumpy, forming well-clustered (highly fractal) stellar distributions. This clustering behavior becomes 
somewhat scattered (larger $D_2$) for older stars, stabilizing at $D_2 \sim$\, 1.4\,-\,1.6, up to $\sim$\,300\,Myr. 
This fractal dimension, however, is still characteristic of self-similarity, 
indicating that while stellar clustering does evolve with time, the original structure in M\,31 persists over more than $\sim$\,300\,Myr. 
This time-scale is in agreement with the lower limits placed for structure survival in dwarf galaxies \citep{bastian11}. 
Values of $D_{2} \sim$\,1.5, measured in the distributions of size and luminosity of star-forming regions in the 
spiral galaxy NGC\,628, and across the whole extend of star-forming galaxies of various types, are interpreted as 
indication of hierarchy induced by turbulence \citep[][]{elmegreen06, 2elmegreen14}. Similarly, we conclude that the observed 
fractal dimensions in all three areas may be driven by turbulent processes on galactic scales.

The change of the TPCF index at $r \simeq S_{\rm break}$ and the stability of $D_2$ to an almost constant value for separations 
$r > S_{\rm break}$ suggests that possibly there is a maximum length-scale up to which stars sustain their clustering pattern, 
i.e., up to which their structures survive. Moreover, the apparent dependence of $S_{\rm break}$ to stellar age implies that this 
scale differs between stars at different evolutionary stages. Although the physical meaning of this break is not yet understood, its 
presence might be related to the scale of the galactic disk. For example, a break in the power spectra of ISM emission in the Large 
Magellanic Cloud is interpreted as due to the line-of-sight thickness of the galactic disk \citep[][]{block10}. Indeed, in spiral galaxies stars 
tend to group up to scales comparable to the disk scale height, and larger structures become spiral-like (flocculent) because of their 
longer dynamical time-scales in comparison to the shear time  \cite[see, e.g.,][]{E2011EAS31}. 
\begin{svgraybox}
{\sl Under these circumstances, with the use of the TPCF, we not only determine the time-scale for structure survival, 
but we may also specify the upper length-scale limit across the galactic disk where structure survives, 
before it `dissolves' in it}. 
\end{svgraybox}

The observed difference in the behavior of the TPCF from one area to other, suggests a possible
dependence of the clustering behavior of stars to the position of the areas across the disk of M\,31, and thus the dynamical
influence of the galaxy. Area\,9 coincides with the first arm at distance $\sim$\,5\,kpc, area\,15 is located at the 10\,kpc star-forming ring, and the more remote 
area\,21 at a distance $\sim$\,15\,kpc away from the galactic centre (see Fig.\,\ref{fig:footprints}, right panel).  
Area\,9, being closer to the centre, experiences stronger gravitational instabilities and possible disruption by the galactic 
potential. This is portrayed by the TPCF (at small separations) and $S_{\rm break}$ variabilities, which  are seen in Fig.\,\ref{fig:tpcf_evol} 
only for area\,9, and not for areas\,15 and 21. 

Considering the rotation of the galaxy \citep[e.g.,][]{sofuerubin01}, based on their positions, 
all three areas have rotational velocities quite similar to each other varying between 236\,km\,s$^{-1}$ at $\sim$\,5\,kpc,  260\,km\,s$^{-1}$ at $\sim$\,10\,kpc, 
and 251\,km\,s$^{-1}$ at $\sim$\,15\,kpc \citep[see, e.g., Table\,1 in][]{carignan06}. While all areas have similar velocities, being at different distances 
from the galactic centre, perform a complete orbit at different time-scales, the closest being fastest. As a consequence, the oldest stars in our samples, 
of $\sim$\,300\,Myr, have performed thus far 10\% of an orbit in area\,9, 6\% in area\,15 and 4\% in area\,21. This suggests that these stars have experienced
different degrees of streaming motions driven by the spiral arms, which scatter stars as they shear by \citep[e.g.,][]{elmegreenstruck13}, 
with those in area\,21 being less disrupted.

Based on the discussion above, the findings of this study can be summarized to the following points: 
\begin{enumerate}

\item Stellar clustering occurs at length-scales that depend on both stellar age and position on the disk.

\item Star formation produces clumpy structures of young stars, on which galactic dynamics influence over time subsequent clustering processes or substructures.

\item Young stellar structures survive across the whole extend of M\,31 for at least $\sim$\,300\,Myr. 

\item The distribution of young stars evolves as a function of stellar age, but remains fractal with $D_2 \sim$\,1.4\,-\,1.6 up to this age limit.
 
\item The observed self-similarity in the stellar distribution is probably induced by large-scale turbulence.

\item Stars in the outer parts away from the galactic centre experience less disruption driven by the gravity of spiral arms.

\end{enumerate}

\begin{acknowledgement}
Based on observations 
made with the NASA/ESA {\sl Hubble 
Space Telescope}, obtained from the data archive at the Space Telescope Science Institute (STScI). STScI is operated by the 
Association of Universities for Research in Astronomy, Inc.\ under NASA contract NAS 5-26555. 
D.A.G. acknowledges support by the German Research Foundation through grant GO\,1659/3-1. 
I also acknowledge very useful discussions with Bruce Elmegreen, Ralf Klessen, Sacha Hony and 
Lukas Konstandin, which helped improve this manuscript. Finally, I would like to thank 
the organizers for a very interesting meeting in a wonderful location, and wish to our honourees 
and their beloved all the best for good health and happiness!  
\end{acknowledgement}

\end{document}